\UseRawInputEncoding

\documentclass[aps,prl,amsmath,amssymb,floatfix,reprint,citeautoscript,noeprint,superscriptaddress,twocolumn]{revtex4-2}
\usepackage{xr}

\usepackage{dsfont}
\usepackage{lipsum} 
\usepackage{bibentry}
\usepackage[english]{babel}
\usepackage[dvipsnames]{xcolor}
\usepackage{graphicx}
\usepackage[caption=false]{subfig} 
\usepackage{amsmath,amssymb,bm}
\usepackage[version=3]{mhchem}
\usepackage{verbatim}
\usepackage{multirow}
\usepackage{dcolumn}
\usepackage{float}
\usepackage{ragged2e}
\usepackage{nicefrac}
\usepackage{siunitx}
\usepackage{booktabs}
\usepackage{chemformula}
\usepackage{wrapfig}
\usepackage{enumitem}  
\usepackage{transparent}
\usepackage[colorlinks,allcolors=black,citecolor=blue,urlcolor=blue]{hyperref}
\usepackage{tcolorbox}
\usepackage{relsize}

\usepackage{sansmathfonts}
\makeatletter
\input{si-labels.aux}
\long\def\@makecaption#1#2{%
  \vskip\abovecaptionskip
  \sffamily 
  \small    
  \begingroup
    \leftskip=0pt plus 1fil
    \rightskip=0pt plus 1fil
    \parfillskip=0pt
    \justifying 
    \noindent
    \bfseries #1: \normalfont #2\par
  \endgroup
  \vskip\belowcaptionskip
}
\makeatother

\usepackage{xr}
\makeatletter
\newcommand*{\addFileDependency}[1]{
\typeout{(#1)}
\@addtofilelist{#1}
\IfFileExists{#1}{}{\typeout{No file #1.}}
}\makeatother

\newcommand*{\myexternaldocument}[1]{%
\externaldocument[si-]{#1}%
\addFileDependency{#1.tex}%
\addFileDependency{#1.aux}%
}
\myexternaldocument{si}

\begin{document}

\def\mytitle{
How reproducible are first-principles simulations of liquid water?
}
\title{\mytitle}
\author{Niamh O'Neill}%
\email{oneilln@mpip-mainz.mpg.de}
\thanks{These authors contributed equally to this work.}
\affiliation{Max Planck Institute for Polymer Research, Mainz 55128, Germany}

\author{Benjamin X. Shi}%
\email{mail@benjaminshi.com}
\thanks{These authors contributed equally to this work.}
\affiliation{Initiative for Computational Catalysis, Flatiron Institute, 160 5th Avenue, New York, NY 10010, USA}

\author{William J. Baldwin}%
\affiliation{%
Lennard-Jones Centre, University of Cambridge, Trinity Ln, Cambridge, CB2 1TN, UK
}
\affiliation{%
Department of Engineering, University of Cambridge, Cambridge, CB3 0HE, UK
}

\author{Albert P. Bart\'ok}%
\affiliation{Department of Physics, University of Warwick, Coventry, CV4 7AL, UK}%
\affiliation{%
Warwick Centre for Predictive Modelling, School of Engineering, University of Warwick, Coventry, CV4 7AL, UK
}

\author{Chris J. Pickard}%
\affiliation{Department of Materials Science and Metallurgy, University of Cambridge, 27 Charles Babbage Road, Cambridge CB3 0FS, UK}%
\affiliation{%
Lennard-Jones Centre, University of Cambridge, Trinity Ln, Cambridge, CB2 1TN, UK
}
\affiliation{Advanced Institute for Materials Research, Tohoku University, Sendai 980-8577, Japan}

\author{Angelos Michaelides}%
\affiliation{Yusuf Hamied Department of Chemistry, University of Cambridge, Lensfield Road, Cambridge CB2 1EW, UK}%
\affiliation{%
Lennard-Jones Centre, University of Cambridge, Trinity Ln, Cambridge, CB2 1TN, UK
}

\author{G\'abor Cs\'anyi}
\affiliation{Max Planck Institute for Polymer Research, Mainz 55128, Germany}
\affiliation{%
Lennard-Jones Centre, University of Cambridge, Trinity Ln, Cambridge, CB2 1TN, UK
}
\affiliation{%
Department of Engineering, University of Cambridge, Cambridge, CB3 0HE, UK
}
\author{Timothy C. Berkelbach}
\affiliation{Initiative for Computational Catalysis, Flatiron Institute, 160 5th Avenue, New York, NY 10010, USA}
\affiliation{Department of Chemistry, Columbia University, New York, NY 10027, USA}%

\begin{abstract}
Liquid water is fundamentally important, and its accurate computer simulation has been the driving force for myriad methodological developments.
\textit{Ab initio} molecular dynamics with forces obtained from density functional theory (DFT) is now a standard tool widely used by researchers.
However, we reveal that previous studies of liquid water using the same widely-used density functional (revPBE-D3) exhibit significant discrepancies with one another, varying by over 20\% in the diffusion coefficient and 10\% in the density, raising fundamental questions about reproducibility.
By combining modern long-range machine-learning interatomic potentials that enable robust statistical sampling with carefully converged DFT training data, we resolve these discrepancies, achieving consensus across six diverse community codes.
Our predictions differ markedly from previous literature: we show that most previous results overestimate the density and underestimate the diffusion coefficient of revPBE-D3 water due to basis set incompleteness and pseudopotential inconsistencies, coupled with limitations in statistical sampling (in some cases).
These benchmark values provide a reliable reference for validating current and future implementations of DFT-based \textit{ab initio} molecular dynamics.
Reaching agreement establishes confidence and credibility and serves as a prerequisite for the systematic assessment of new density functionals and numerical approximations.

\end{abstract}

{\maketitle}

\section{Introduction}

Density functional theory (DFT) is a central computational tool in the natural and engineering sciences, driving progress in applications ranging from drug design to materials discovery.
Over the years, its popularity and utility has given rise to an extensive set of computer codes capable of predicting the energies, atomic forces, and other properties of complex systems.
However, in all codes, the DFT equations are only solved approximately, and different codes implement different approximations.
Ideally, the predictions of one should be reproducible by another, but the extent to which this is realized depends on the differences in their underlying implementation details.
This understanding has motivated major community-driven efforts that have highlighted the difficulty in achieving reproducibility across DFT codes~\cite{lejaeghereReproducibilityDensityFunctional2016,bosoniHowVerifyPrecision2024}.
Similar efforts have also emerged for first-principles methods beyond DFT~\cite{vansettenGW100BenchmarkingG0W02015,
dellapiaReproducibilityFixednodeDiffusion2025}, which are often even more sensitive to numerical approximations.

These large-scale efforts have provided valuable insights that make reproducibility across codes now readily achievable for the \SI{0}{\kelvin} properties of solids. 
Unfortunately, comparable efforts have not yet been extended to the properties of liquids, which have even greater reproducibility challenges.
Specifically, computing liquid properties is significantly more expensive, requiring \textit{ab initio} molecular dynamics (AIMD) that involves hundreds of thousands of DFT evaluations.
This added cost discourages the use of tight numerical settings in the underlying DFT and pragmatic choices must often be made.
In addition to the DFT approximations, the need to achieve statistical convergence is an added dimension, and insufficient sampling may further preclude reproducibility. 

\begin{figure*}[t]
    \includegraphics[width=1.0\textwidth]{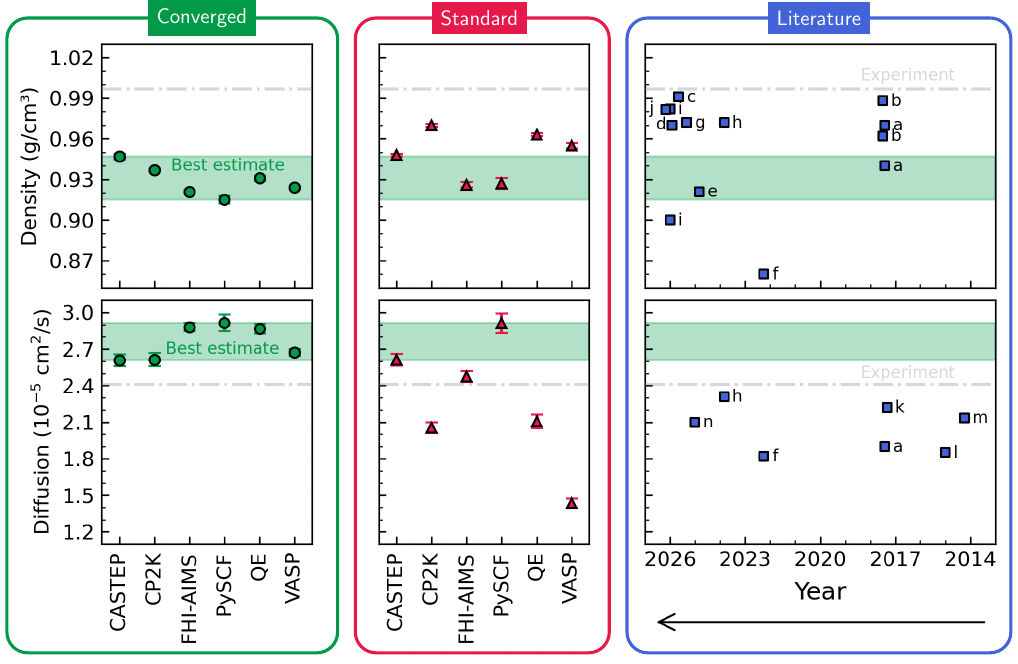}
    \caption{%
    \textbf{Reproducibility of revPBE-D3 for density and diffusion coefficient of water at ambient conditions.}
    Left and middle panels (green and red): Density (top) and self-diffusion coefficient (bottom) from MLIPs trained to revPBE-D3 using converged and standard numerical settings, respectively, from 6 independent codes given in the x-axis.
    Diffusion coefficient values have been corrected for finite-size effects with the Yeh-Hummer approach~\cite{yehSystemsizeDependenceDiffusion2004}.
    Sensitivity to the choice of thermostat is given in Table~\ref{si-tab:nve} in the SI.
    Each point corresponds to the average of 6 independent replicas over 6 model seeds (each run for at least 1 ns), and error bars depict the standard error.
    Right panel (blue): Literature values for density (top) and self-diffusion coefficient (bottom) for revPBE-D3 as a function of the year published. Some studies feature multiple estimates corresponding to different electronic structure parameters.
    The labels correspond to literature as follows: a.\cite{pestanaInitioMolecularDynamics2017}, b.\cite{galibMassDensityFluctuations2017}, c.\cite{brookesCO2HydrationAir2025}, d.\cite{ferrettiAccurateSimulationsWater2025}, 
    e.\cite{monterodehijesDensityIsobarWater2024}, 
    f.\cite{zaverkinPredictingPropertiesPeriodic2022}, g.\cite{zhaoNeuralNetworkBasedMolecular2025}, h.\cite{avulaUnderstandingAnomalousDiffusion2023}, i.\cite{liInitioMeltingProperties2025}, j.\cite{limRevealingStrainEffects2026}, k.\cite{marsalekQuantumDynamicsSpectroscopy2017}, l.\cite{bankuraStructureDynamicsSpectral2014},
    m.\cite{dingAnomalousWaterDiffusion2014},
n.\cite{ibrahimEfficientParametrizationTransferable2024}. We discuss and tabulate all the values in the SI, Sec.~\ref{si-sec:lit_review}.}
\label{fig:main}
\end{figure*}

In this work, we consider the reproducibility of simulations of liquid water, which is perhaps the most studied system by AIMD (see e.g. refs.~\cite{gillanPerspectiveHowGood2016,tuckermanInitioSimulationsWater1994,fernandez-serraNetworkEquilibrationFirstprinciples2004,grossmanAssessmentAccuracyDensity2004,schweglerAssessmentAccuracyDensity2004,vandevondeleInfluenceTemperatureDensity2004,sitStaticDynamicalProperties2005,santraCoupledClusterBenchmarks2009,ceriottiNuclearQuantumEffects2016,chenInitioTheoryModeling2017,lacountEnsembleFirstprinciplesMolecular2019,zhangModelingLiquidWater2021,villardStructureDynamicsLiquid2024}).
Specifically, we will focus on simulations with the revPBE functional~\cite{perdewGeneralizedGradientApproximation1996,zhangCommentGeneralizedGradient1998} with D3 dispersion corrections~\cite{grimmeConsistentAccurateInitio2010} (revPBE-D3), which is one of the most popular functionals for liquid water in recent years.
A survey of the literature confirms a reproducibility challenge that is illustrated in the right panel of Figure~\ref{fig:main}, which compiles literature predictions of two fundamental properties of liquid water (with classical nuclei): the density and the diffusion coefficient.
Despite all using the same revPBE-D3 functional, we observe a substantial spread in the density (\SIrange{0.860}{0.991}{\gram\per\cubic\centi\metre}) and the diffusion coefficient (\SIrange{1.82e-5}{2.31e-5}{\square\centi\metre\per\second}).
This variation, which is comparable to the variation between density functionals~\cite{gillanPerspectiveHowGood2016,monterodehijesDensityIsobarWater2024}, can lead to conflicting conclusions about the accuracy of different functionals.
For example, some revPBE-D3 studies report excellent agreement with experiment, while others find poor agreement, with deviations exceeding 10\%.
These variations can lead to to qualitative errors in predicted properties, such as the phase diagram~\cite{chengInitioThermodynamicsLiquid2019,boreRealisticPhaseDiagram2023}.
Ensuring reproducibility across DFT implementations is therefore essential for the reliability of both subsequent scientific studies and industrial applications.

As we will demonstrate in this work, these variations primarily arise from the numerical approximations used in DFT codes---in particular, the treatment of core electrons and accuracy of pseudopotentials, as well as the type and size of the basis set used to represent the DFT orbitals.
Previous studies~\cite{pestanaInitioMolecularDynamics2017,galibMassDensityFluctuations2017,daoSystematicTrendsWater2026,beckMoreConvergedLess2026} have examined some of these numerical choices within individual DFT codes, but reproducibility and verification across different codes remain open questions.
An additional factor is limited statistical sampling to converge the thermodynamic observables, which we find particularly prevalent in the earlier studies cited in Figure~\ref{fig:main} (see SI, Figure~\ref{si-fig:diff_spread}).
In more recent studies, these sampling bottlenecks have been largely alleviated by employing machine-learning interatomic potentials (MLIPs) that reproduce DFT at lower cost~\cite{chengInitioThermodynamicsLiquid2019,boreRealisticPhaseDiagram2023, morawietzHowVanWaals2016, schranMachineLearningPotentials2021, singraberDensityAnomalyWater2018, chengNuclearQuantumEffects2016, kapilInexpensiveModelingQuantum2020, omranpourPerspectiveAtomisticSimulations2024, piaggiPhaseEquilibriumWater2021, zhangPhaseDiagramDeep2021, xuAccuratePredictionHeat2023, riemelmoserMachineLearningDensity2023}.
However, significant discrepancies remain because the MLIPs are ultimately limited by the numerical approximations in the underlying DFT training data.
In addition, their ability to reproduce DFT can be constrained by model architecture~\cite{hilpertAccurateThermophysicalProperties2026} and by incomplete or noisy training data~\cite{kurylaHowAccurateAre2025}.

In this work, we establish precise estimates of the density and diffusion coefficient of liquid water at the revPBE-D3 level with classical nuclei under ambient conditions.
For completeness and comparison with experiments, we also report these properties from path integral molecular dynamics simulations (i.e., with quantum nuclei).
We show that standard numerical settings, which control the underlying approximations, are typically insufficient to converge these quantities, but with tight settings, agreement can be achieved across six DFT codes.
Such tight settings are rarely used to perform AIMD because of their high computational cost, which is a challenge that we overcome, like in recent work, by training a MLIP that  allows exhaustive statistical sampling at an affordable cost.
Our converged density and diffusion coefficient differ substantially from most previous works and serve as new reference benchmarks for future work.
Such benchmarks are a necessary prerequisite for assessing the accuracy of existing functionals or developing new functionals.

\section{Results and Discussion}
\subsection{A reliable machine-learning protocol}
To overcome the sampling challenge of AIMD with precise numerical settings, we train an MLIP for each of six DFT codes with tightly converged settings.
For each of these MLIPs, properties are computed by averaging over multiple trajectories from different model seeds, resulting in cumulative simulations of several nanoseconds in length.
Training data are generated following recently validated protocols~\cite{monterodehijesDensityIsobarWater2024,oneillRoutineCondensedPhase2025}, yielding an extensive set of configurations spanning multiple pressures.
We employ the recently introduced local split-charge MACE architecture~\cite{parkerFalseMetallizationShortranged2026}, which augments conventional short-range MACE with global long-range electrostatic interactions (See SI, Sec.~\ref{si-sec:mlip} for full details of dataset and model).
The resulting model achieves state-of-the-art fidelity to DFT, reducing root-mean-square errors on the forces to $\sim$\SI{4}{meV\per\AA}, which is a factor of two smaller than short-range MACE (SI, Tables \ref{si-tab:sr_lr_errors} and \ref{si-tab:sr_lr_obs}) and one to two orders of magnitude smaller than early-generation MLIPs~\cite{monterodehijesDensityIsobarWater2024,liInitioMeltingProperties2025}.
Because of their low force errors, our MLIPs reproduce the corresponding AIMD densities to within \SI{0.01}{\gram\per\cubic\centi\metre} (SI, Sec.~\ref{si-sec:aimd}).

The six DFT codes (listed alphabetically) are selected to span a range of numerical design philosophies: CASTEP~\cite{clarkFirstPrinciplesMethods2005} (ultrasoft pseudopotentials with plane-wave basis sets), CP2K~\cite{kuhneCP2KElectronicStructure2020} (pseudopotentials with Gaussian-type orbitals and plane-wave auxiliary basis sets), FHI-aims~\cite{blumInitioMolecularSimulations2009} (all-electron with numerical atomic orbitals), PySCF~\cite{sunPySCFPythonbasedSimulations2018} (all-electron with Gaussian-type orbitals), Quantum ESPRESSO~\cite{giannozziQUANTUMESPRESSOModular2009} (projector-augmented-wave method or pseudopotentials with plane-waves), and VASP~\cite{kresseInitioMolecularDynamics1993a,kresseEfficiencyAbinitioTotal1996,kresseEfficientIterativeSchemes1996} (projector-augmented-wave method with plane-wave basis sets).
For each code, we use the best available pseudopotentials (or all-electron treatments where possible), and we employ tightly converged numerical settings to approach the basis-set limit.
For example, in plane-wave codes we use energy cutoffs more than four times higher than standard recommendations (SI, Table~\ref{si-tab:standard_converged}), while for Gaussian-type orbitals we employ quadruple-$\zeta$ basis sets augmented with diffuse functions.
In addition to performing periodic calculations with five codes, we also perform molecular calculations with PySCF, and we train the MLIP using a recently developed $\Delta$-learning framework~\cite{oneillRoutineCondensedPhase2025} that bridges gas-phase cluster calculations to condensed-phase periodic descriptions (see SI, Sec.~\ref{si-sec:delta}).

\subsection{Best estimates to the density and diffusion}

The left panel of Figure~\ref{fig:main} shows the density and diffusion coefficients predicted by MLIPs trained on data generated with each DFT code under tightly converged numerical settings.
The predictions are consistent across all six codes, falling within a narrow range of $0.930 \pm 0.017\,\si{\gram\per\cubic\centi\metre}$ for the density and $(2.75 \pm 0.15) \times\SI{e-5}{\square\centi\metre\per\second}$ for the diffusion coefficient.
The predicted density is lower than most previous reports, and the diffusion coefficient is larger than all previous reports (Figure~\ref{fig:main} right panel).
Using one of our tightly converged MLIPs, we perform path integral MD simulations to account for nuclear quantum effects (see SI, Sec.~\ref{si-sec:nqe}), from which we predict a density of $0.927 \pm 0.001\,\si{\gram\per\cubic\centi\metre}$ and diffusion coefficient of $(2.55 \pm 0.07)\times\SI{e-5}{\square\centi\metre\per\second}$.
Compared to the experimental values of \SI{0.997}{\gram\per\cubic\centi\metre} and \SI{2.41e-5}{\square\centi\metre\per\second}~\cite{holzTemperaturedependentSelfdiffusionCoefficients2000}, the predicted density is a significant underestimation, but the diffusion coefficient is reasonably accurate.

The observed agreement and reproducibility across all DFT codes is made possible by the combination of tightly converged numerical settings and advanced MLIP architectures.
Our employed settings are significantly more stringent than the `standard' ones that are commonly used, recommended, or set by default---for example, using softer pseudopotentials and smaller basis sets (see Methods and SI, Sec.~\ref{si-sec:settings}).
The middle panel of Figure~\ref{fig:main} shows density and diffusion coefficient predictions from MLIPs trained on data generated with such standard settings.
The predictions exhibit a broad spread, ranging from \SIrange{0.926}{0.970}{\gram\per\cubic\centi\metre} for the density and \SIrange{1.44e-5}{2.91e-5}{\square\centi\metre\per\second} for the diffusion coefficient, consistent with the spread seen in previous reports.

\subsection{Rationalizing discrepancies}

The discrepancies with previous literature can be largely attributed to pseudopotential errors and basis set incompleteness.
Due to the popularity of the CP2K code, most studies employ Goedecker-Teter-Hutter (GTH) pseudopotentials~\cite{goedeckerSeparableDualspaceGaussian1996} parameterized for PBE together with a triple-$\zeta$ (TZV2P) basis set, yielding densities between \SIrange{0.962}{0.991}{\gram\per\cubic\centi\metre} and diffusion coefficients between \SIrange{1.85e-5}{2.31e-5}{\square\centi\metre\per\second}.
These numerical settings were selected for our `standard' CP2K setup (Figure~\ref{fig:main}, middle panel), and the associated predictions (\SI{0.970}{\gram\per\cubic\centi\metre}, \SI{2.05e-5}{\square\centi\metre\per\second}) are consistent with previous studies.
As shown in the SI, Sec.~\ref{si-sec:pyscf_gth}, the resulting deviations from our converged best estimates (${\sim}\SI{0.04}{\gram\per\cubic\centi\metre}$ in density and ${\sim}\SI{0.6e-5}{\square\centi\metre\per\second}$ in diffusion coefficient) are primarily due to the basis set: when increasing the basis from TZV2P to aug-cc-GRB-Q (quadruple-$\zeta$ with diffuse functions), these deviations are reduced by ${\sim}80\,$\%.

Replacing the GTH-PBE pseudopotential with the consistent GTH-revPBE pseudopotential eliminates the remaining discrepancy, bringing both properties into agreement with our best estimates.
The errors from using an inconsistent GTH-PBE pseudopotential can also be reproduced with PySCF (SI, Sec.~\ref{si-sec:pyscf_gth}), which can perform all-electron and pseudopotential calculations on equal footing.
We note that the effect of using the inconsistent GTH-PBE pseudopotential does not appear for static $0\,$K benchmarks such as the WATER27 dataset nor the radial distribution function (SI, Sec.~\ref{si-sec:rdf}).
In other codes, we typically find it necessary to use harder pseudopotentials than are standard and concomitant larger energy cutoffs, or larger atomic orbital basis sets augmented by diffuse functions.

Finally, we highlight that there exist several variants of the D3 dispersion correction that must be distinguished.
The original formulation~\cite{grimmeConsistentAccurateInitio2010} employs zero damping, denoted D3(0), and this is the variant used throughout this work.
Alternative forms include Becke-Johnson damping~\cite{beckeDensityfunctionalModelDispersion2005} [D3(BJ)], and a three-body Axilrod-Teller-Muto (ATM) dispersion term~\cite{axilrodInteractionVanWaals1943} may be optionally combined with either damping scheme.
For completeness, we report densities and diffusion coefficients for these alternative variants in the SI, Table~\ref{si-tab:atm_density_diffusion}. 
Notably, we observe differences of up to \SI{0.04}{\gram\per\cubic\centi\metre} in the density and more than \SI{1.0e-5}{\square\centi\metre\per\second} in the diffusion coefficient, underscoring the subtle yet significant influence of dispersion~\cite{morawietzHowVanWaals2016} on molecular liquids such as water.
These results further highlight the importance of explicitly specifying the chosen D3 variant, as the default may vary across DFT codes.

\section{Conclusions}

To conclude, our work demonstrates that high-precision and reproducible first-principle simulations of liquids are now attainable.
We show that tight numerical convergence, coupled with exhaustive sampling afforded by MLIPs are necessary conditions for such reproducibility.
We achieve agreement across six diverse DFT codes for the density and diffusion coefficient of liquid water using the revPBE-D3 density functional approximation, yielding values within a narrow range of $0.930 \pm 0.017,\si{\gram\per\cubic\centi\metre}$ for the density and $(2.75 \pm 0.15) \times \SI{e-5}{\square\centi\metre\per\second}$ for the diffusion coefficient.
This contrasts sharply with prior literature, where we show that incomplete basis sets, inconsistent pseudopotentials and insufficient sampling have led to significant discrepancies.

This work has important implications for both developers and practitioners of density functional theory.
For developers, our work motivates the design and implementation of improved approximations controlling basis set incompleteness and pseudopotential errors.
Our best estimates reported here for revPBE-D3 water can serve as reference benchmarks to validate such efforts.
Importantly, we expect the impact of these approximations to be even more severe for electronic structure theories beyond the generalized gradient approximation, such as hybrid or double-hybrid functionals and wavefunction-based theories.
For practitioners, our work establishes guidelines for the numerical settings required to reliably study complex systems with DFT, thereby increasing confidence and building credibility in reported results while helping to mitigate future reproducibility challenges.

As a final corollary, our work shows that ``standard'' settings in many codes are insufficient to reach quantitatively converged, and thereby reliable, DFT predictions.
Although pragmatic concerns have previously demanded that such standard numerical settings be used in AIMD, this limitation can be overcome using modern MLIPs, which can now closely reproduce the underlying DFT with tightly converged settings.
Even with such converged DFT training data, it is important to use expressive MLIP architectures with well-curated datasets, as it has recently been shown that errors from simpler MLIP architectures introduce notable deviations in predictions for the physical properties of liquid water~\cite{hilpertAccurateThermophysicalProperties2026}.
Crucially, the cost of MLIP training and inference is independent of the level of convergence of the training data, so the high one-time cost of generating such reference data is worthwhile.
This framework enables routine simulations that can be considered ``exact'' AIMD, potentially obviating the need to perform AIMD ever again for anything other than generation of training data.

\section{Methods}
\subsection{Density functional theory}

We distinguish between between ``standard'' and more stringent ``converged'' settings that we describe in more detail in the SI, Sec.~\ref{si-sec:settings}.
The converged calculations employ larger basis sets (e.g., verytight in FHI-aims~\cite{blumInitioMolecularSimulations2009}, def2-QZVPPD in PySCF~\cite{sunPySCFPythonbasedSimulations2018,sunRecentDevelopmentsPySCF2020}, aug-ccGRB-Q in CP2K~\cite{kuhneCP2KElectronicStructure2020}), and harder or more accurate pseudopotentials (or none, when possible).
Further, (integration) grid densities and numerical precision settings are also tightened in each code.
Standard settings use more moderate basis sets (e.g., intermediate in FHI-aims, def2-TZVPD in PySCF, TZV2P in CP2K), lower energy cutoffs (e.g., 520 eV in VASP~\cite{kresseInitioMolecularDynamics1993a,kresseEfficiencyAbinitioTotal1996,kresseEfficientIterativeSchemes1996} and 84 Ry in Quantum Espresso~\cite{giannozziQUANTUMESPRESSOModular2009}), and default or recommended pseudopotentials.
Relatively large energy cutoffs were required even in the standard settings of CP2K ($1200\,$Ry) and CASTEP ($1500\,$eV) to sufficiently reduce noise in the forces~\cite{kurylaHowAccurateAre2025} for MLIP training.
As all periodic configurations involved 64-waters (${\sim}12\,$\AA{} in unit cell length), we employ only a $\Gamma$-point sampling of the Brillouin zone.
Additionally, we added the D3 with zero-damping and without three-body dispersion terms using the \texttt{simple-dftd3} Python package~\cite{ehlertSimpleDFTD3Library2024}.

\subsection{Machine-learning interatomic potential}

We trained machine-learning interatomic potential (MLIP) models  for each DFT code.
The models for CASTEP, CP2K, FHI-aims, Quantum ESPRESSO, and VASP were trained on periodic datasets of 64-water configurations generated from MD simulations across a wide pressure range using an intermediate revPBE-D3 MACE model.
The final dataset contained 1591 structures (10\% taken out for validation), each labeled with energies, forces, and stresses.
We employed a MACE architecture with two message-passing layers (128 channels, $6\,$\AA{} cutoff), augmented with a local split-charge scheme~\cite{parkerFalseMetallizationShortranged2026} to capture electrostatics, yielding energy errors below $0.2\,$meV/atom and force errors below $7\,$meV/\AA{}.
For the PySCF model, we employ a recently proposed $\Delta$-learning framework~\cite{oneillRoutineCondensedPhase2025}, training a smaller secondary MACE model (2 layers, 64 channels, $4\,$\AA{} cutoff) to learn the difference between converged FHI-aims and PySCF energies from molecular clusters, achieving errors of $\sim$0.2 meV/atom.

\subsection{Molecular dynamics}

All molecular dynamics (MD) simulations were carried out using the Atomic Simulation Environment~\cite{hjorthlarsenAtomicSimulationEnvironment2017} with a $0.5\,$fs timestep at $298\,$K.
NPT simulations employed a Nos\'e-Hoover-type barostat at 1 bar with a $500\,$fs relaxation time, and densities were obtained via the mean and standard error over six independent trajectories from different model seeds of at least 400 ps each.
A 64 water box was used to replicate previous simulation setups.
Diffusion coefficients were computed from NVT simulations at the experimental density using 64 water molecules in a fixed $12.42\,$\AA{} box, with a CSVR thermostat ($50\,$fs equilibration followed by ${\geq}700\,$ps production run with a $1\,$ps relaxation time).
Additional NVE simulations were performed to quantify thermostat effects.
Diffusion constants were extracted from the slope of the mean squared displacement averaged over six seeds, and corrected for finite-size effects using the Yeh-Hummer approach~\cite{yehSystemsizeDependenceDiffusion2004}, with remaining size errors expected to be small.

\section*{Data availability}
All data required to reproduce the findings of this study will be made available upon publication of this study.

\section*{Code availability}
All simulations were performed with publicly available simulation software (\texttt{ACEsuit}, \texttt{ASE}).

\section*{Acknowledgments}

We thank Matthias Baumann, Kara Fong, Yair Litman, Thomas Markland, Xavier Rosas Advincula and Christoph Schran for helpful discussions.
N.O.N. acknowledges support from the Gates Cambridge Foundation, as well as the hospitality of the Initiative for Computational Catalysis at the Flatiron Institute, where a portion of this work was carried out. 
The Flatiron Institute is a division of the Simons Foundation. 
A.M. acknowledges support from the European Union under the ``n-AQUA'' European Research Council project (Grant No.\ 101071937). 

\bibliography{references.bib}

\end{document}


\title{\mytitle}

\date{\today}

\author{Niamh O'Neill}%
\email{oneilln@mpip-mainz.mpg.de}
\thanks{These authors contributed equally to this work.}
\affiliation{Max Planck Institute for Polymer Research, Mainz 55128, Germany}

\author{Benjamin X. Shi}%
\email{mail@benjaminshi.com}
\thanks{These authors contributed equally to this work.}
\affiliation{Initiative for Computational Catalysis, Flatiron Institute, 160 5th Avenue, New York, NY 10010, USA}

\author{William J. Baldwin}%
\affiliation{%
Lennard-Jones Centre, University of Cambridge, Trinity Ln, Cambridge, CB2 1TN, UK
}
\affiliation{%
Department of Engineering, University of Cambridge, Cambridge, CB3 0HE, UK
}

\author{Albert P. Bart\'ok}%
\affiliation{Department of Physics, University of Warwick, Coventry, CV4 7AL, UK}%
\affiliation{%
Warwick Centre for Predictive Modelling, School of Engineering, University of Warwick, Coventry, CV4 7AL, UK
}

\author{Chris J. Pickard}%
\affiliation{Department of Materials Science and Metallurgy, University of Cambridge, 27 Charles Babbage Road, Cambridge CB3 0FS, UK}%
\affiliation{%
Lennard-Jones Centre, University of Cambridge, Trinity Ln, Cambridge, CB2 1TN, UK
}
\affiliation{Advanced Institute for Materials Research, Tohoku University, Sendai 980-8577, Japan}

\author{Angelos Michaelides}%
\affiliation{Yusuf Hamied Department of Chemistry, University of Cambridge, Lensfield Road, Cambridge CB2 1EW, UK}%
\affiliation{%
Lennard-Jones Centre, University of Cambridge, Trinity Ln, Cambridge, CB2 1TN, UK
}

\author{G\'abor Cs\'anyi}
\affiliation{Max Planck Institute for Polymer Research, Mainz 55128, Germany}
\affiliation{%
Lennard-Jones Centre, University of Cambridge, Trinity Ln, Cambridge, CB2 1TN, UK
}
\affiliation{%
Department of Engineering, University of Cambridge, Cambridge, CB3 0HE, UK
}
\author{Timothy C. Berkelbach}
\affiliation{Initiative for Computational Catalysis, Flatiron Institute, 160 5th Avenue, New York, NY 10010, USA}
\affiliation{Department of Chemistry, Columbia University, New York, NY 10027, USA}%

\maketitle
\tableofcontents

\newpage

\section{\label{sec:protocol}Protocol for reliable simulations of liquids}
We will describe and summarize the protocol we have developed to ensure reliable and reproducible simulations of liquids here.
We will start by discussing the architecture and dataset of our machine-learning interatomic potential (MLIP), which achieves state-of-the-art performance.
Then, we will discuss further details on how we performed molecular dynamics with these MLIPs to compute the density and diffusion coefficient of liquid water.

\subsection{\label{sec:mlip}Machine learning interatomic potential}

The MLIP models based on CASTEP, CP2K, FHI-aims, Quantum ESPRESSO and VASP were trained on periodic reference data labelled with energies, forces and stresses. 
Following a similar approach to our \cite{oneillRoutineCondensedPhase2025} and others' \cite{monterodehijesDensityIsobarWater2024} previous work, a robust dataset for computing densities was generated by sampling molecular dynamics (MD) trajectories of 64-water boxes generated at a range of pressures (specifically -1500, -500, -300, 1, 2, 500, 1000, 4000, 8000 bar).
We used a previously developed revPBE-D3 MLIP~\cite{oneillRoutineCondensedPhase2025} based on a short-range MACE architecture to generate configurations at the appropriate level of theory.
The final dataset comprised 1591 structures, with 10\% held out as a validation set.
The dataset and models will be made available on Github on publication of the paper.

Our MLIP models utilize a MACE architecture~\cite{batatiaMACEHigherOrder2022} that has been recently augmented~\cite{parkerFalseMetallizationShortranged2026} to include explicit electrostatic interactions based on partial atomic charges (learned directly from energies and gradients).
This local split charge MACE model ensures charge conservation by redistributing the charge between neighboring atoms (initially set to their formal oxidation states of +2 and -1 for oxygen and hydrogen atoms respectively) based on the chemical environment.
We note that in the overall design space of electrostatic models, this is a simple model with locally determined charges.
While other more sophisticated models are available \cite{baldwinDesignSpaceSelfConsistent2026}, for the purposes of this work, the local-split-charge models are sufficient.

Overall, the MACE architecture specifically comprised 2 message passing layers with 128 channels and a 6 \AA{} cutoff.
Total energies, forces and stresses were included in the loss, with initial weights 100, 100, 10000 respectively.
The energy weight was then increased in a second stage to 1000.
In all cases, six seeds were trained, and the final simulation results are averaged over all seeds, with the standard error reported as error bars.
Validation errors are given in Table \ref{tab:errors} for the five periodic DFT codes, split into the `converged' and `standard' numerical settings.
The resulting fits are generally excellent, with energy errors below $0.2\,$meV/atom and force errors below $7\,$meV/\AA{} for the converged DFT codes.
This error is a bit worse for the standard codes, due to the increased numerical noise, leading to energy errors of $0.8\,$meV/atom in some cases, although force errors remain low.
We note that to achieve such low force errors (and thereby reliable MLIPs) for the standard codes, larger cutoffs than is typical were required.

\begin{table}[h!]
\caption{Validation root-mean-square-errors for the MLIPs trained to the converged and standard electronic structure settings (described in more detail in Section \ref{sec:settings}) below.}
\label{tab:errors}
\begin{tabular}{@{}lrrrrrr@{}}
\toprule
             & \multicolumn{3}{c}{Converged}                                                                                                                                                                  & \multicolumn{3}{c}{Standard}                                                                                                                                                                  \\ \cmidrule(lr){2-4} \cmidrule(lr){5-7}
Code & \shortstack{Energy \\ (meV/atom)}
& \shortstack{Force \\ (meV/\AA{})}
& \shortstack{Stress \\ (meV/\AA{}$^3$)}
& \shortstack{Energy \\ (meV/atom)}
& \shortstack{Force \\ (meV/\AA{})}
& \shortstack{Stress \\ (meV/\AA{}$^3$)} \\  \midrule
CASTEP           & 0.08                                                             & 6.88                                                        & 0.13                                                          & 0.09                                                            & 5.63                                                        & 0.07                                                          \\
CP2K             & 0.17                                                             & 5.81                                                        & 0.11                                                          & 0.09                                                            & 5.78                                                       & 0.13                                                          \\
FHI-aims         & 0.07                                                             & 3.82                                                        & 0.07                                                          & 0.08                                                            & 5.17                                                        & 0.10                                                          \\
Quantum Espresso & 0.08                                                             & 4.89                                                        & 0.07                                                          & 0.29                                                            & 5.25                                                        & 0.09                                                          \\
VASP             & 0.06                                                             & 3.99                                                        & 0.06                                                          & 0.79                                                            & 5.85                                                        & 0.17                                                          \\ \bottomrule
\end{tabular}
\end{table}

In Table~\ref{tab:sr_lr_errors}, we compare the effect of using the long-range local split charge MACE model against a purely local short-range MACE model when using converged FHI-aims data.
The improvements are notable in both the energy and forces, with the latter more than halving.
We notice a minor deterioration in the stress errors but these remain small.
The errors for both models are overall very low, and they predict densities and diffusion coefficients which are in excellent agreement with one another in Table~\ref{tab:sr_lr_obs}.

\begin{table}[h!]
\caption{Validation errors for the local split charge long-range MACE model and standard short-range MACE model trained to the FHI-aims converged dataset.}
\label{tab:sr_lr_errors}

\begin{tabular}{@{}lccc@{}}
\toprule
            & \begin{tabular}[c]{@{}c@{}}Energy \\ {[}meV/atom{]}\end{tabular} & \begin{tabular}[c]{@{}c@{}}Force\\ {[}meV/\AA{}{]}\end{tabular} & \begin{tabular}[c]{@{}c@{}}Stress\\ {[}meV/\AA{}$^3${]}\end{tabular} \\ \midrule
Long-range  & 0.07                                                             & 3.82                                                        & 0.07                                                          \\
Short-range & 0.10                                                             & 7.72                                                        & 0.03                                                          \\ \bottomrule
\end{tabular}
\end{table}

\begin{table}[h!]
\centering
\caption{Density and diffusion coefficients from short-range and local split-charge long-range MACE models trained on FHI-aims with converged settings dataset.}
\label{tab:sr_lr_obs}
\begin{tabular}{@{}lcc@{}}
\toprule
            & Density (g/cm$^{3}$) & Diffusion (10$^{-5}$ cm$^2$/s) \\ \midrule
Long-range  & 0.921 $\pm$ 0.001      & 2.87 $\pm$ 0.06                 \\
Short-range & 0.920 $\pm$ 0.001      & 2.78 $\pm$ 0.07                 \\ \bottomrule
\end{tabular}
\end{table}

\subsection{\label{sec:delta}Delta-learning MLIP}

For the PySCF code, we performed gas-phase calculations to access large highly-converged basis sets.
We use a recent $\Delta$-learning framework developed by some of us~\cite{oneillRoutineCondensedPhase2025} to perform reliable condensed phase simulations when just trained to gas-phase cluster data.
This framework involves both a baseline model trained to periodic data and $\Delta$-model trained to molecular clusters.
For the baseline model, we used the long-range converged FHI-aims model developed in Section~\ref{sec:mlip}.
The $\Delta$-model is trained to the difference between FHI-aims and PySCF.
Thus, when the baseline and $\Delta$ models are summed together, we are able to obtain the properties calculated using PySCF.
We have demonstrated that this approach works highly effectively in Ref.~\citenum{oneillRoutineCondensedPhase2025}, where the $\Delta$-framework was used to correct from one density functional approximation (DFA) to another as well as up to coupled cluster theory [CCSD(T)].

The molecular cluster dataset used to train the $\Delta$-model were taken from trajectories of molecular dynamics simulations of liquid water (against across several pressures) --- the same set of configurations used in Ref. \cite{oneillRoutineCondensedPhase2025}.
The short-range $\Delta$-model comprising 2 message-passing layers, 64 channels and a 4 \AA\ cutoff (justification for this smaller model size than the periodic baseline is given in Ref. \cite{oneillRoutineCondensedPhase2025}) was trained on the difference between the FHI-aims and PySCF energies.
We developed $\Delta$-models for PySCF with tightly converged and default standard settings.
In addition to all-electron calculations shown in Figure~1 of the main text, we also trained models using the GTH-PBE pseudopotentials as described in Section~\ref{sec:pyscf_gth}.
Energy validation errors were at ${\sim}0.2\,$meV/atom for all of the models.

\clearpage

\subsection{Molecular dynamics computational details}
All molecular dynamics simulations were performed using the Atomic Simulation Environment \cite{hjorthlarsenAtomicSimulationEnvironment2017} Python package, with a timestep of 0.5 fs at 298 K.
A box size of 64 waters was used to be consistent with most previous literature.
Pressure was controlled in NPT simulations using a Nos\'e-Hoover-type barostat, with an external isotropic pressure of 1 bar applied using a barostat relaxation time of 500 fs.
The mean density was computed from the block averaged simulations from six seeds of at least 400 ps length each.
Diffusion coefficients were computed from simulations in the cannonical (NVT) ensemble.
For consistent comparison with previous literature, NVT simulations were done at the experimental density with 64 waters and a fixed box size of 12.42 \AA{}.
The canonical sampling through velocity rescaling (CSVR) thermostat~\cite{bussiCanonicalSamplingVelocity2007} was used, and a short NVT simulation was first performed with a 50 fs relaxation time, followed by production simulations with a 1 ps temperature relaxation time for at least 700 ps.
To check the dependance of the diffusion coefficient on the choice of thermostat (parameters), for the FHI-aims short-range model shown in Table \ref{tab:sr_lr_obs}, we also computed the diffusion coefficient from NVE simulations.
50 replicas were initialised from a NVT simulation at 300 K, and simulation propagated using the velocity verlet dyanmics in ASE, with a timestep of 0.5 fs.
Comparison is given in Table~\ref{tab:nve} below.

The self-diffusion coefficient was obtained from fitting the slope of the mean squared displacement of the water oxygen atoms versus time from 0.1 to 10 ps. 
Each result is the average and standard error of 6 independent model seeds.
The diffusion coefficient obtained for the 64 water simulation box $D(L)$ was corrected for the finite size effects from using a periodic simulation cell of length $L$ using the Yeh and Hummer correction \cite{yehSystemSizeDependenceDiffusion2004}:
\begin{equation}
    D(L) = D(\infty) - \zeta\frac{k_BT}{6\pi \eta L}
\end{equation}
where $\eta$ was taken as the experimental sheer viscosity of 0.8925 mPa$\cdot$s and the periodic self-interaction energy for a cubic simulation cell $\zeta$ as 2.837297.
We note that there may be remaining finite-size effects arising from using a 64 water box (used here for consistency with previous literature), however previous tests~\cite{morawietzHowVanWaals2016} suggest this is small ($\sim 0.1$~cm$^2$/s on the density).

\begin{table}[h!]
\caption{Comparison of diffusion coefficient computed for the FHI-aims converged model from NVT simulations with CSVR thermostat and NVE simulations.}
\label{tab:nve}
\begin{tabular}{@{}lc@{}}
\toprule
           & Diffusion (10$^{-5}$ cm$^2$/s) \\ \midrule
NVT (CSVR) & 2.879 +/- 0.034                \\
NVE        & 2.843 +/- 0.080                \\ \bottomrule
\end{tabular}
\end{table}

\begin{figure}[h]
    \includegraphics[width=0.8\textwidth]{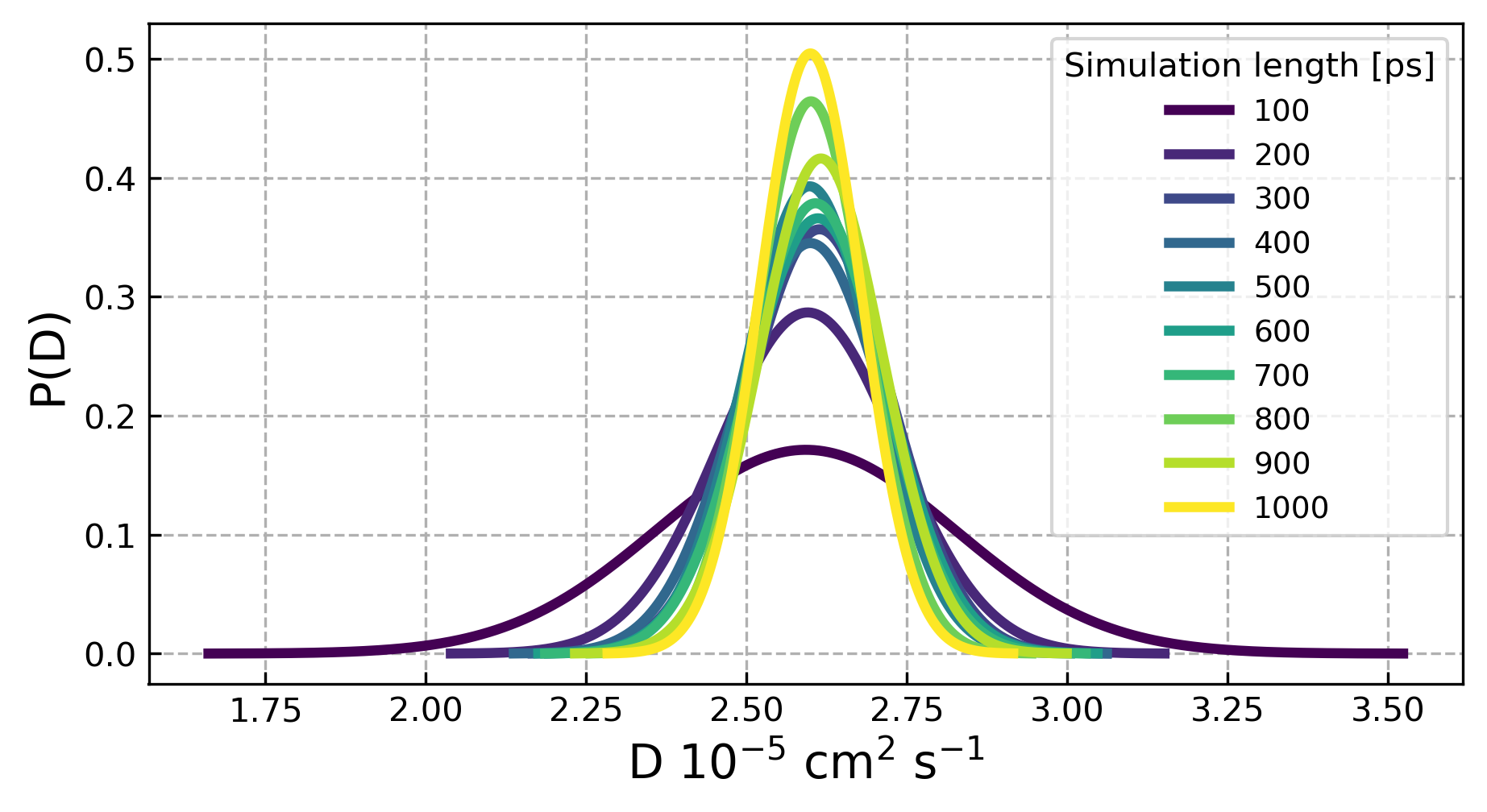}
    \caption{%
    Distribution of calculated diffusion coefficients as a function of simulation time. 
    }
    \label{fig:diff_spread}
\end{figure}

Additionally Figure \ref{fig:diff_spread} compares the spread in predicted diffusion coefficients as a function of simulation time, highlighting the importance of sufficiently long trajectories to obtain reliable statistics.
Notably, simulation lengths below $100\,$ps (seen in several earlier AIMD-based studies in Table~\ref{tab:water_diffusion}) can lead to estimates covering a $95\,$\% confidence range of over \SI{0.9e-5}{\square\centi\metre\per\second} when reporting only a single simulation.

\clearpage

\section{Validation against \textit{ab initio} MD}\label{sec:aimd}
While MLIPs can largely alleviate the sampling bottleneck highlighted in the Introduction of the main text, to obtain reliable thermodynamic observables, the question remains whether the underlying model can itself reproduce the ground truth.
The most rigorous test is to compare to \textit{ab initio} simulations.
We performed \textit{ab initio} molecular dynamics simulations using Quantum ESPRESSO, and compared the prediction of the density to that of the MLIP trained to reproduce Quantum ESPRESSO.
A total of 12 replicas were initialized from independent initial configurations and run for at least 100 ps each.
The results are summarized in Table \ref{tab:aimd} below, where the AIMD and MACE model predictions agree to within 0.012 g/cm$^{3}$.
\begin{table}[h!]
\centering
\caption{Density from AIMD simulations based on Quantum ESPRESSO and the long-range MACE MLIP trained to Quantum ESPRESSO.}
\label{tab:aimd}
\begin{tabular}{lc}
\toprule
 & Density (g/cm$^{3}$)  \\
\midrule
AIMD & $0.943 \pm 0.002$ \\
MACE & $0.931 \pm 0.002$ \\
\bottomrule
\end{tabular}
\end{table}

\clearpage

\section{\label{sec:settings}Electronic structure details}

We discuss the code-specific electronic structure parameters in this section.
They are summarized in Table~\ref{tab:standard_converged}, where we provide settings for both `standard' settings (either commonly used in the literature or the defaults recommended by developers) and the tight `converged' settings we have used to reach the high precision needed for accurate densities and diffusion coefficients.
The key differences lie in the choice of basis set quality (controlled by plane-wave cutoff or basis set size) and the inclusion of a pseudopotential.
All periodic calculations used $\Gamma$-point sampling of the Brillouin zone.

\begin{table}[h]
\caption{Summary of standard and converged settings used in the training of the MACE models.}
\label{tab:standard_converged}
\begin{tabular}{lll}
\toprule
Code             & Type      & Settings  \\ \midrule
\multirow{2}{*}{CASTEP}             & Standard  & 1500 eV cutoff and PBE `C19mk2' pseudopotentials (PPs)   \\
                 & Converged & 3000 eV cutoff and revPBE `HARD` PPs \\ \midrule
\multirow{2}{*}{CP2K}             & Standard  & 1200 Ry cutoff, TZV2P-GTH basis set and GTH-PBE PPs   \\
                 & Converged & 2400 Ry cutoff, aug-ccGRB-Q basis set and GTH-revPBE PPs \\ \midrule

\multirow{2}{*}{FHI-aims}         & Standard  & `intermediate' basis set\\
                 & Converged & `verytight' basis set \\ \midrule
\multirow{2}{*}{PySCF} & Standard  & def2-TZVPD basis set \\
                 & Converged & def2-QZVPPD basis set \\ \midrule
\multirow{2}{*}{\parbox[t]{2.5cm}{\RaggedRight
Quantum\\
Espresso
}} & Standard  & 84 Ry wavefunction cutoff, ONCV PPs \\
                 & Converged & 100 (800) Ry wavefunction (density) cutoff, SSSP `Precision' PPs                                       \\ \midrule
\multirow{2}{*}{VASP}             & Standard  & 520 eV cutoff, default VASP PAW potentials \\
                 & Converged & 3000 eV cutoff, hard (GW-optimized) VASP PAW potentials \\

                 \bottomrule
\end{tabular}
\end{table}

\clearpage
\subsection{FHI-aims}

DFT calculations from the Fritz Haber Institute \textit{ab initio} molecular simulations (FHI-aims) package~\cite{blumInitioMolecularSimulations2009} were performed on version 250320.
For the `converged' settings, we employed `verytight' basis sets for H and O atoms, while this was decreased to `intermediate' for the `standard' settings.
Scalar relativistic effects were included in both calculations.

\subsection{VASP}

DFT calculations from the Vienna \textit{ab-initio} simulation package~\cite{kresseInitioMolecularDynamics1993a,kresseEfficiencyAbinitioTotal1996,kresseEfficientIterativeSchemes1996} (VASP) were performed on version 6.5.1.
The `converged' settings employed a large $3000\,$eV energy cutoff coupled with the GW-optimized hard (denoted by \texttt{\_h\_GW}) projector augmented wave (PAW) potentials.
We also used dense ({\tt PREC=Accurate}) grids for these calculations.
The `standard' settings employed a $520\,$eV energy cutoff coupled with the standard PAW potentials and {\tt PREC=Normal} grids.

\subsection{Quantum Espresso}

Our Quantum Espresso calculations~\cite{giannozziQUANTUMESPRESSOModular2009} were performed on version 7.5.
The `converged' settings employed pseudopotentials from the \texttt{Precision} Standard Solid State Pseudopotentials~\cite{prandiniPrecisionEfficiencySolidstate2018}, together with a wavefunction energy cutoff of $100\,$Ry and charge density cutoff of $800\,$Ry.
The `standard' settings employed the PseudoDojo norm-conserving pseudopotentials~\cite{vansettenPseudoDojoTrainingGrading2018} with the recommended energy cutoff of $84\,$Ry and the density cutoff at 4$\times$ this value.

\subsection{CP2K}

CP2K~\cite{kuhneCP2KElectronicStructure2020} calculations were performed on v2025.2.
The `converged' settings utilized the aug-ccGRB-Q basis sets with the GTH-revPBE~\cite{goedeckerSeparableDualspaceGaussian1996} pseudopotentials.
A plane-wave cutoff with $2400\,$Ry was used for the auxiliary plane-wave basis.
The `standard' settings utilized the TZV2P basis set with a $1200\,$Ry auxiliary plane-wave basis.

\subsection{PySCF}

We performed all-electron (AE) calculations on PySCF~\cite{sunPySCFPythonbasedSimulations2018,sunRecentDevelopmentsPySCF2020}.
The `standard' calculations utilized the def2-TZVPD basis set.
The `converged' calculations uses the def2-QZVPPD basis set.
We also performed GTH-PBE pseudopotential calculations in Section~\ref{sec:pyscf_gth} using the aug-QZV2P basis set.

\subsection{CASTEP}

CASTEP~\cite{clarkFirstPrinciplesMethods2005} calculations were performed on version 25.12.
The `standard' calculations utilized the default `C19mk2' ultrasoft pseudopotentials, with the $1500\,$eV plane-wave cutoff.
The `converged' calculation moves to the `HARD' pseudopotentials, coupled wtih an energy cutoff of $3000\,$eV.

\subsection{D3}

The D3 calculations were performed separately on the Simple DFT-D3 package~\cite{ehlertSimpleDFTD3Library2024}.
The energies, forces and stresses were subsequently summed with each of the DFT calculations.

\clearpage

\section{Final density and diffusion estimates}

The left and middle panels of Figure 1 of the main text graphically illustrates the final predicted diffusion and density predictions for the `converged' and `standard' numerical settings, respectively.
We tabulate these final estimates for the density and diffusion coefficient in Tables~\ref{tab:density} and~\ref{tab:diffusion}, respectively.

\begin{table}[h]
\caption{Comparison of the density predictions of liquid water for standard and converged numerical settings across the six DFT codes.}
\label{tab:density}
\begin{tabular}{lrr}
\toprule
Code & Standard (g/cm$^{3}$)  & Converged (g/cm$^{3}$) \\
\midrule
FHI-aims & 0.926 $\pm$ 0.002 & 0.921 $\pm$ 0.001 \\
VASP & 0.955 $\pm$ 0.002 & 0.924 $\pm$ 0.001 \\
Quantum Espresso & 0.963 $\pm$ 0.001 & 0.931 $\pm$ 0.002 \\
CP2K & 0.970 $\pm$ 0.001 & 0.937 $\pm$ 0.001 \\
PySCF & 0.927 $\pm$ 0.004 & 0.915 $\pm$ 0.003 \\
CASTEP & 0.948 $\pm$ 0.001 & 0.947 $\pm$ 0.002 \\
\bottomrule
\end{tabular}
\end{table}

\begin{table}[h]
\caption{Comparison of the predictions of the diffusion coefficients for standard and converged numerical settings across the six DFT codes.}
\label{tab:diffusion}
\begin{tabular}{lrr}
\toprule
Code & Standard (10$^{-5}$ cm$^2$/s) & Converged (10$^{-5}$ cm$^2$/s) \\
\midrule
FHI-aims & 2.474 $\pm$ 0.046 & 2.879 $\pm$ 0.034 \\
VASP & 1.435 $\pm$ 0.039 & 2.672 $\pm$ 0.030 \\
Quantum Espresso & 2.184 $\pm$ 0.02 & 2.867 $\pm$ 0.041 \\
CP2K & 2.054 $\pm$ 0.046 & 2.613 $\pm$ 0.053 \\
PySCF & 2.911 $\pm$ 0.080 & 2.914 $\pm$ 0.069 \\
CASTEP & 2.611 $\pm$ 0.048 & 2.607 $\pm$ 0.049 \\
\bottomrule
\end{tabular}
\end{table}

\subsection{\label{sec:nqe}Nuclear quantum effects}
For completeness we have also computed the diffusion coefficient and radial distribution function (at the experimental density) and the density via path integral molecular dynamics (PIMD) simulations to incorporate nuclear quantum effects (NQEs).
We used the the short-range FHI-aims model (which we show in Table \ref{tab:sr_lr_obs} gives the same results as the long-range model), which could be used in i-PI \cite{litmanIPI30Flexible2024} simulations interfaced with LAMMPS \cite{thompsonLAMMPSFlexibleSimulation2022} and Symmetrix \cite{wcwittWcwittSymmetrix2026, kovacsEvaluationMACEForce2023, batatiaFoundationModelAtomistic2025} for efficient PIMD simulations.
PIMD simulations were performed using 32 beads with a 0.25 fs timestep at 298 K in a simulation box with 64 waters.
The density was computed from simulations in the NPT ensemble, using a Langevin thermostat with $\tau=100$ fs, with the PILE thermostat used for the barostat with $\tau=10$ fs. 
Each simulation for the density was 250 ps long, with the average and standard error taken over 12 replicas.
Thermostatted-RPMD (T-RPMD) \cite{rossiHowRemoveSpurious2014} simulations were performed in the canonical (NVT) ensemble to obtain the diffusion coefficient.
A PILE thermostat with $\lambda=0.5$ and $\tau=100$ fs was used.
The mean squared displacement of the centroid of the ring polymer was used to compute the diffusion coefficient, with the final value being the average and standard error over 10 replicas each 100 ps long.
We find that the diffusion coefficient decreases upon the inclusion of NQEs (Table \ref{tab:nqes}), in agreement with previous work \cite{marsalekQuantumDynamicsSpectroscopy2017}.
However, unlike the previous work (using the GTH-PBE PP and TZV2P basis set) which saw poorer experimental agreement upon inclusion of NQEs, here we show that inclusion of NQEs actually brings revPBE-D3 into closer agreement with the experimental value of 2.41.
Meanwhile, there is a slight increase in the density upon inclusion of NQEs.

\begin{table}[h!]
\centering
\caption{Density and diffusion coefficient from PIMD and molecular dynamics simulations with classical nuclei.}
\label{tab:nqes}
\begin{tabular}{@{}lcc@{}}
\toprule
            & Density (g/cm$^{3}$) & Diffusion (10$^{-5}$ cm$^2$/s) \\ \midrule
Quantum Nuclei (PIMD)  & 0.927 $\pm$ 0.001      & $2.549 \pm 0.07$                 \\
Classical Nuclei (MD) & 0.921 $\pm$ 0.001       & $2.776 \pm 0.07$                \\ \bottomrule
\end{tabular}
\end{table}
\begin{figure}[h]
    \includegraphics[width=1.0\textwidth]{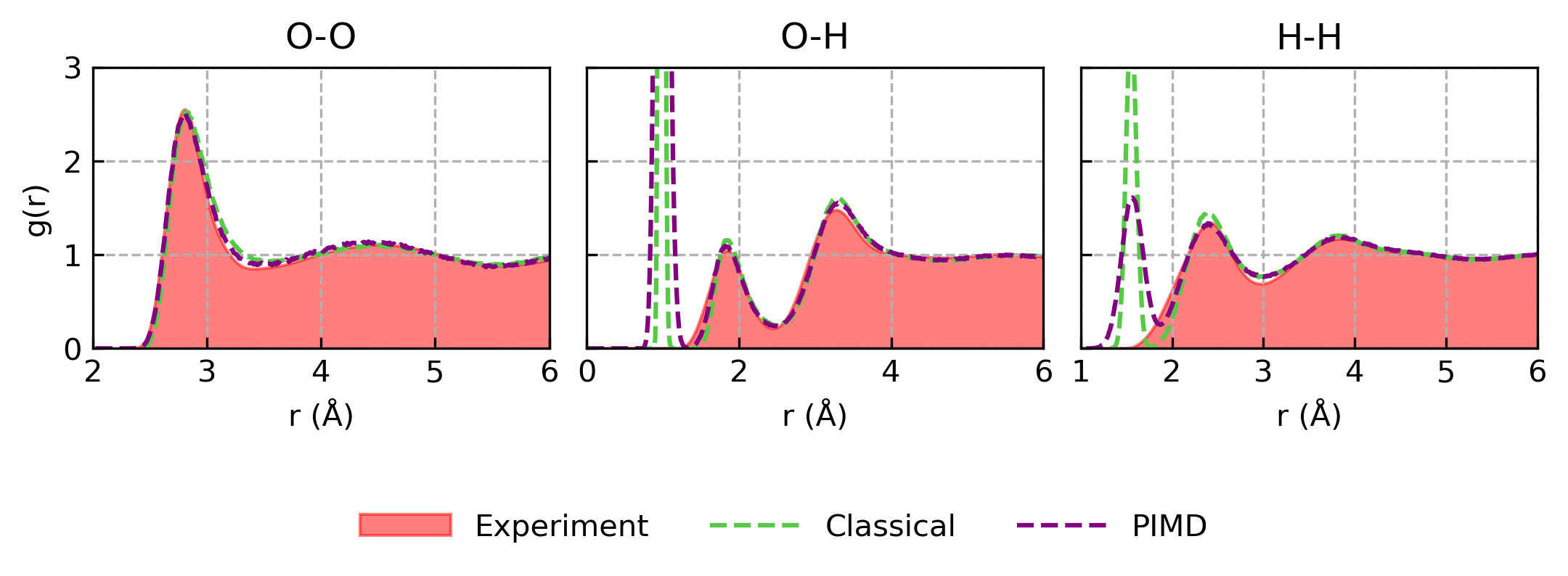}
    \caption{%
    Comparison of O-O, O-H and H-H RDFs from PIMD and classical molecular dynamics simulations. 
    }
    \label{fig:rdfs}
\end{figure}

\clearpage
\section{\label{sec:lit_review}Literature compilation}
The right panel of Figure 1 of the main text graphically illustrates the range in predicted diffusion and density predictions over the years, by various different codes and approaches.
Tables \ref{tab:water_density} and \ref{tab:water_diffusion} below summarise the technical details of each work, including the code used, method (\textit{ab initio} or MLIP based), plane-wave cutoff and basis set size (where relevant).

\begin{table}[h!]
\caption{Summary of literature values for the density (reported in g/cm$^3$) of liquid water at 298-300 K.}
\label{tab:water_density}
\footnotesize
\begin{tabular}{lccccccc}
\toprule
Year & Density & Code & Method & Basis & PW cutoff & Pseudopotential & Ref. \\
\midrule
2017 & 0.970 & CP2K & AIMD & mTZV2P & 800 Ry & GTH-PBE~\cite{goedeckerSeparableDualspaceGaussian1996} &\cite{pestanaInitioMolecularDynamics2017} \\
2017 & 0.940 & CP2K & AIMD & mTZV2P & 400 Ry & GTH-PBE~\cite{goedeckerSeparableDualspaceGaussian1996} & \cite{pestanaInitioMolecularDynamics2017} \\
2017 & 0.962 & CP2K & AIMD & TZV2P & 800 Ry & GTH-PBE~\cite{goedeckerSeparableDualspaceGaussian1996} & \cite{galibMassDensityFluctuations2017} \\
2017 & 0.988 & CP2K & AIMD & MOLOPT-DZVP-SR & 800 Ry & GTH-PBE~\cite{goedeckerSeparableDualspaceGaussian1996} & \cite{galibMassDensityFluctuations2017} \\
2022 & 0.860 & Turbomole & GM-NN~\cite{zaverkinGaussianMomentsPhysically2020} & def2-TZVP & - & AE &\cite{zaverkinPredictingPropertiesPeriodic2022} \\
2023 & 0.972 & CP2K & DPMD~\cite{wangDeePMDkitDeepLearning2018} & TZV2P & 1200 Ry & GTH-PBE~\cite{goedeckerSeparableDualspaceGaussian1996} & \cite{avulaUnderstandingAnomalousDiffusion2023} \\
2024 & 0.921 & VASP & BP-NNP~\cite{behlerGeneralizedNeuralNetworkRepresentation2007} & PW & 2000 eV & Hard PAW & \cite{monterodehijesDensityIsobarWater2024} \\
2025 & 0.991 & CP2K & MACE~\cite{batatiaMACEHigherOrder2022} & TZV2P & 650 Ry & GTH-PBE~\cite{goedeckerSeparableDualspaceGaussian1996} & \cite{brookesCO2HydrationAir2025} \\
2025 & 0.970 & Quantum Espresso & MACE~\cite{batatiaMACEHigherOrder2022} & PW & 60 Ry & pslibrary.1.0.0~\cite{dalcorsoPseudopotentialsPeriodicTable2014} & \cite{ferrettiAccurateSimulationsWater2025} \\
2025 & 0.972 & CP2K & DPMD~\cite{wangDeePMDkitDeepLearning2018} & DZVP & N/A & GTH-PBE~\cite{goedeckerSeparableDualspaceGaussian1996} & \cite{zhaoNeuralNetworkBasedMolecular2025} \\
2025 & 0.919 & Quantum Espresso & DPMD~\cite{wangDeePMDkitDeepLearning2018} & PW & 110 Ry & ONCV~\cite{schlipfOptimizationAlgorithmGeneration2015} & \cite{liInitioMeltingProperties2025} \\
2026 & 0.982 & CP2K & MACE~\cite{batatiaMACEHigherOrder2022} & TZV2P & 1050 Ry & GTH-PBE~\cite{goedeckerSeparableDualspaceGaussian1996} & \cite{limRevealingStrainEffects2026} \\
\bottomrule
\end{tabular}
\end{table}

\begin{table}
\caption{Summary of previous literature values for liquid water self-diffusion coefficient (reported in 10$^{-9}$ m$^2$/s) at 298-300$\,$K. }
\label{tab:water_diffusion}
\footnotesize
\begin{tabular}{lcccccccc}
\toprule
Year & Diffusion & Code & Method & Basis & PW cutoff & Pseudopotential & Length [ps] & Ref. \\
\midrule
2014 & 1.850 & CP2K & AIMD & TZV2P & 400 Ry &  GTH-PBE~\cite{goedeckerSeparableDualspaceGaussian1996} & 40 &  \cite{bankuraStructureDynamicsSpectral2014} \\
2014 & 2.135 & CP2K & AIMD & DZV2P & 280 Ry & GTH-PBE~\cite{goedeckerSeparableDualspaceGaussian1996} & 170 &  \cite{dingAnomalousWaterDiffusion2014} \\
2017 & 1.900 & CP2K & AIMD & mTZV2P & 800 Ry & GTH-PBE~\cite{goedeckerSeparableDualspaceGaussian1996} & 40 &  \cite{pestanaInitioMolecularDynamics2017} \\
2017 & 2.220 & CP2K & AIMD & TZV2P & 400 Ry & GTH-PBE~\cite{goedeckerSeparableDualspaceGaussian1996} & 4x200 &  \cite{marsalekQuantumDynamicsSpectroscopy2017} \\
2022 & 1.820 & Turbomole & GM-NN & def2-TZV2P & - & AE & 10x200  & \cite{zaverkinPredictingPropertiesPeriodic2022} \\
2023 & 2.310 & CP2K & DPMD & TZV2P & 1200 Ry & GTH-PBE~\cite{goedeckerSeparableDualspaceGaussian1996} & 16x500 & \cite{avulaUnderstandingAnomalousDiffusion2023} \\
2024 & 2.100 & VASP & ACE & - & 450 Ry & Standard PAW & 100x100 &\cite{ibrahimEfficientParametrizationTransferable2024} \\
\bottomrule
\end{tabular}
\end{table}

\clearpage

\section{Effect of D3 dispersion variants on density and diffusion}
While the standard choice of D3 dispersion is zero-damping, with 2-body terms, in Table \ref{tab:atm_density_diffusion} we also compare the density and diffusion coefficient of the other major damping scheme -- Becke-Johnson, along with the Axilrod-Teller-Muto term to capture 3-body terms in the dispersion.

\begin{table}[h!]
\caption{Effect of the Axilrod--Teller--Muto (ATM)~\cite{axilrodInteractionVanWaals1943} term on the computed density and self-diffusion coefficient for Becke-Johnson and zero-damping schemes.}
\label{tab:atm_density_diffusion}
\centering
\begin{tabular}{@{}lllcc@{}}
\toprule
Damping scheme & ATM? &
Density (g\,cm$^{-3}$) &
Diffusion ($10^{-5}$ cm$^{2}$\,s$^{-1}$) \\ \midrule
\multirow{2}{*}{Zero damping}
 & Yes & $0.901 \pm 0.002$ & $2.512 \pm 0.046$ \\
 & No   & $0.921 \pm 0.001$ & $2.820 \pm 0.048$ \\ \midrule
\multirow{2}{*}{Becke-Johnson}
 & Yes & $0.882 \pm 0.001$ & $1.479 \pm 0.032$ \\
 & No   & $0.895 \pm 0.001$ & $2.672 \pm 0.038$ \\ \bottomrule
\end{tabular}

\end{table}

\clearpage

\section{\label{sec:pyscf_gth}Basis set and pseudopotential errors}

As discussed in the main text, and shown in Tables \ref{tab:water_density} and \ref{tab:water_diffusion} in Section \ref{sec:lit_review}, the majority of the previous literature employ the TZV2P basis set and GTH-PBE pseudopotential.
This raises questions on the origins of their inaccuracies and we will show in this section it comes from both basis set incompleteness of the TZV2P basis set and the pseudopotential inconsistencies from using the GTH-PBE for the revPBE(-D3) DFA.

\subsection{Basis set incompleteness of the TZV2P-GTH basis set}

\begin{figure}[h!]
    \centering
    \includegraphics[width=1\linewidth]{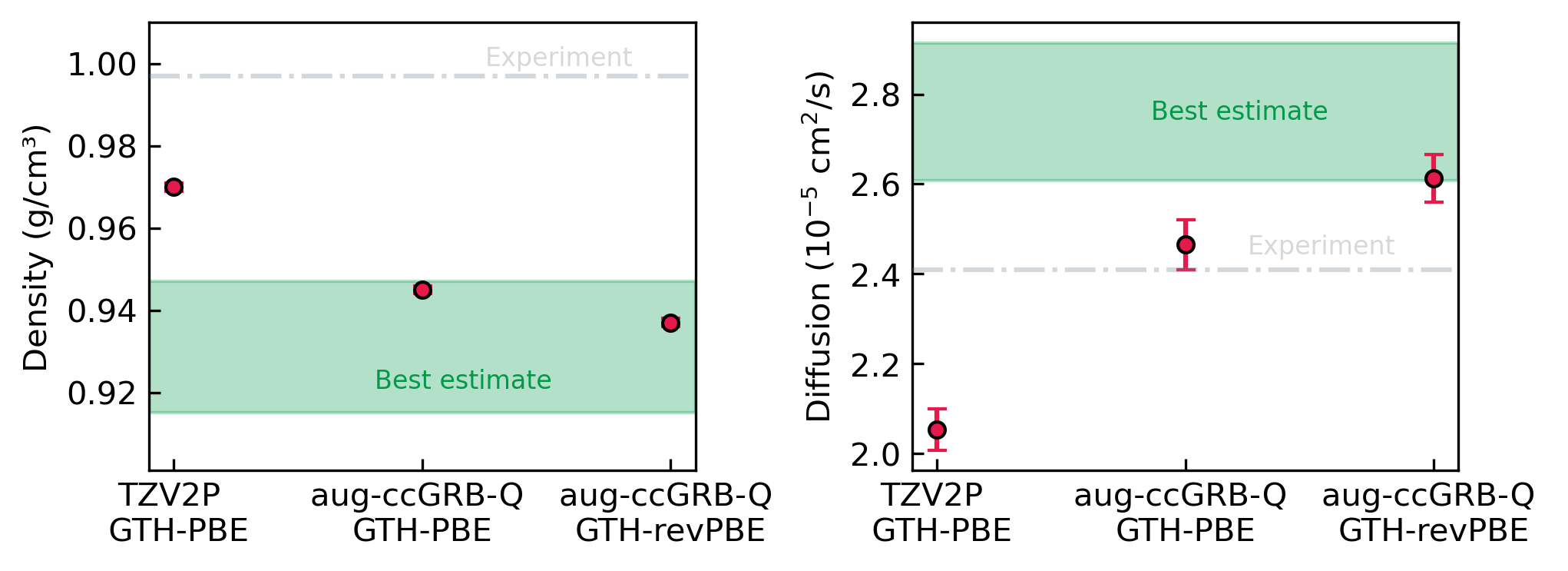}
    \caption{Effect of basis set and pseudopotential on the density and diffusion coefficient of revPBE-D3 liquid water with CP2K.}
    \label{fig:basis_pp}
\end{figure}

In CP2K, it is possible to move beyond the triple-$\zeta$ TZV2P-GTH basis set towards even larger basis sets.
Specifically we use the quadruple-$\zeta$ ccGRB-Q basis set for the tightly `converged' numerical settings which has been enhanced with the aug-cc-Q augmentation functions (this will be dubbed aug-ccGRB-Q).
We showcase the difference between these two basis set choices in Figure \ref{fig:basis_pp} and tabulated in Table~\ref{tab:pp}.
It can be seen that the effect of increasing the basis set is significant, where it both lowers the density from 0.970 g/cm$^{3}$ to 0.945 g/cm$^{3}$, while also increasing the diffusion coefficient from $2.054\times10^{-5}$ cm$^2$/s to $2.465\times10^{-5}$ cm$^2$/s.
The TZV2P-GTH result in Table~\ref{tab:pp} corresponds to the `standard' settings in Table~\ref{tab:standard_converged} of Section~\ref{sec:settings}.
The aug-ccGRB-Q result in Table~\ref{tab:pp} differs from the `converged' settings only in the pseudopotential, where it continues to use the GTH-PBE pseudopotential developed for the PBE exchange-correlation functional.

\begin{table}[h!]
\centering
\caption{Density and diffusion coefficients for CP2K with increasing basis set and different pseudopotentials.}
\label{tab:pp}
\begin{tabular}{@{}lcc@{}}
\toprule
Basis set / Pseudopotential & Density (g/cm$^{3}$) & Diffusion (10$^{-5}$ cm$^2$/s) \\
\midrule
TZV2P / GTH-PBE          & 0.970 $\pm$ 0.001 & 2.054 $\pm$ 0.045 \\
aug-ccGRB-Q / GTH-PBE    & 0.945 $\pm$ 0.001 & 2.465 $\pm$ 0.055 \\
aug-ccGRB-Q / GTH-revPBE & 0.937 $\pm$ 0.001 & 2.613 $\pm$ 0.053 \\
\bottomrule
\end{tabular}
\end{table}

\subsection{Inconsistency errors in the GTH pseudopotentials}

There are also further errors associated with using the GTH-PBE pseudopotentials for the revPBE(-D3) exchange-correlation functional.
This use of inconsistent pseudopotentials has been common within many of the previous studies (see Section~\ref{sec:lit_review}) due to the ready availability of PBE-optimized ones by default in CP2K.
However, a revPBE-optimized GTH pseudopotential is also available~\cite{OpenCPMDGTHpseudopotentials2024} from the OpenCPMD project.
In Figure \ref{fig:basis_pp} and tabulated in Table~\ref{tab:pp}, we have also compared the difference between the GTH-PBE and GTH-revPBE pseudopotentials, both with the aug-ccGRB-Q basis set.
The pseudopotential inconsistency errors has a significant impact, with density decreasing from 0.945 to 0.937, and diffusion coefficient increasing from 2.465 to 2.613.
Once the GTH-revPBE pseudopotential is used ---  corresponding to the converged numerical settings in Table~\ref{tab:standard_converged} --- the density and diffusion coefficients come into good agreement with the remaining codes.

\begin{figure}[h!]
    \centering
    \includegraphics[width=1\linewidth]{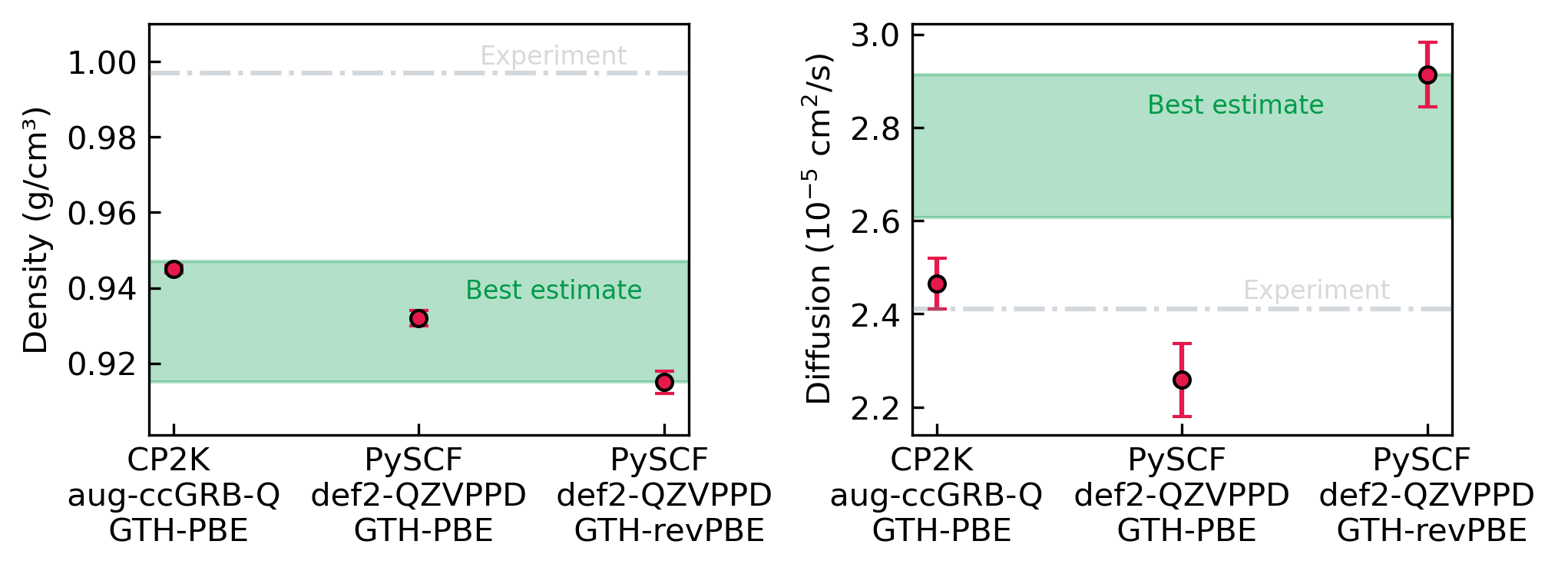}
    \caption{Estimating errors with the GTH-PBE pseudopotential (in the basis set limit) using PySCF.}
    \label{fig:gthpp}
\end{figure}

Separately, it is also possible to use the GTH-PBE pseudopotential in the PySCF code.
This can be used to directly compare against its all-electron calculations (i.e., `converged' numerical settings).
Assuming that we have reached the basis set limit, the only difference would then come from the GTH-PBE pseudopotential.
In Table~\ref{tab:gthpp} and Figure~\ref{fig:gthpp}, we compare the density and diffusion coefficient predicted between the GTH-PBE (with aug-QZV2P basis set) and all-electron (with def2-QZVPPD basis set) estimates.
It can be seen that the GTH-PBE pseudopotential causes an increase in the density and a significant decrease in the diffusion coefficient.
The PySCF GTH-PBE estimates are in good agreement with the CP2K GTH-PBE estimates with the large basis set in Table~\ref{tab:gthpp}.
These results serve to validate that a significant source of errors come from the inconsistent GTH-PBE pseudopotential.

\begin{table}[h!]
\centering
\caption{Tabulated values of the density and diffusion coefficient between all-electron and GTH-PBE pseudopotential calculations in PySCF.}
\label{tab:gthpp}
\begin{tabular}{@{}lcc@{}}
\toprule
Code/ Basis set / Pseudopotential & Density (g/cm$^{3}$) & Diffusion (10$^{-5}$ cm$^2$/s) \\ \midrule
CP2K / aug-ccGRB-Q / GTH-PBE            & 0.977 $\pm$ 0.001      & 1.774 $\pm$ 0.045                \\
PySCF / aug-QZV2P / GTH-PBE     & 0.932 $\pm$ 0.002      & 2.258 $\pm$ 0.078                \\
PySCF / def2-QZVPPD / All-electron    & 0.915 $\pm$ 0.003      & 2.914 $\pm$ 0.069                \\ \bottomrule
\end{tabular}
\end{table}

As a brief aside, we also want to highlight here that such pseudopotential inconsistency errors are also present in CASTEP.
In Table~\ref{tab:castep}, we compare the density and diffusion coefficient between PBE-optimized and revPBE-optimized `HARD' pseudopotentials generated by CASTEP.
It can be seen that there is a slight increase in the diffusion coefficient, which leads to improved overall agreement to the other DFT codes (seen in Figure 1 of the main text).

\begin{table}[h]
\centering
\caption{\label{tab:castep}Comparison of density and self-diffusion coefficient for CASTEP using PBE and revPBE-optimized HARD pseudopotentials.}
\begin{tabular}{lrr}
\toprule
Method &
Density (\si{\gram\per\cubic\centi\metre}) &
Diffusion (10$^{-5}$ cm$^2$/s) \\
\midrule
PBE-optimized HARD & $0.945 \pm 0.001$ & $2.534 \pm 0.078$ \\
revPBE-optimized HARD          & $0.947 \pm 0.002$ & $2.607 \pm 0.049$ \\
\bottomrule
\end{tabular}
\end{table}

\section{\label{sec:rdf}Validation tests}

In this section, we highlight some additional validation tests performed within this study.
First, we will compare radial distribution functions (with classical nuclei) between the different codes --- another important hallmark of liquid water structural parameters.
Subsequently, we will compare the WATER27 dataset of water cluster binding energies to assess the effect of both a smaller basis set (TZV2P) and the effect of the GTH-PBE pseudopotential on water cluster binding energies.

\subsection{Radial distribution function}
In Figure \ref{fig:rdfs} we compare the RDFs predicted by all six DFT codes with both converged and standard numerical settings.
The RDFs with converged numerical settings exhibit extremely subtle changes and are in agreement with the experimentally predicted RDFs from X-ray diffraction \cite{SkinnerStructureWater} and neutron scattering \cite{SoperRadialDistribution2013}.
Meanwhile the RDFs with standard numerical settings show some wider variation (which is well correlated with the extent of the discrepancies in the diffusion coefficent.)

\begin{figure}[h]
    \includegraphics[width=1.0\textwidth]{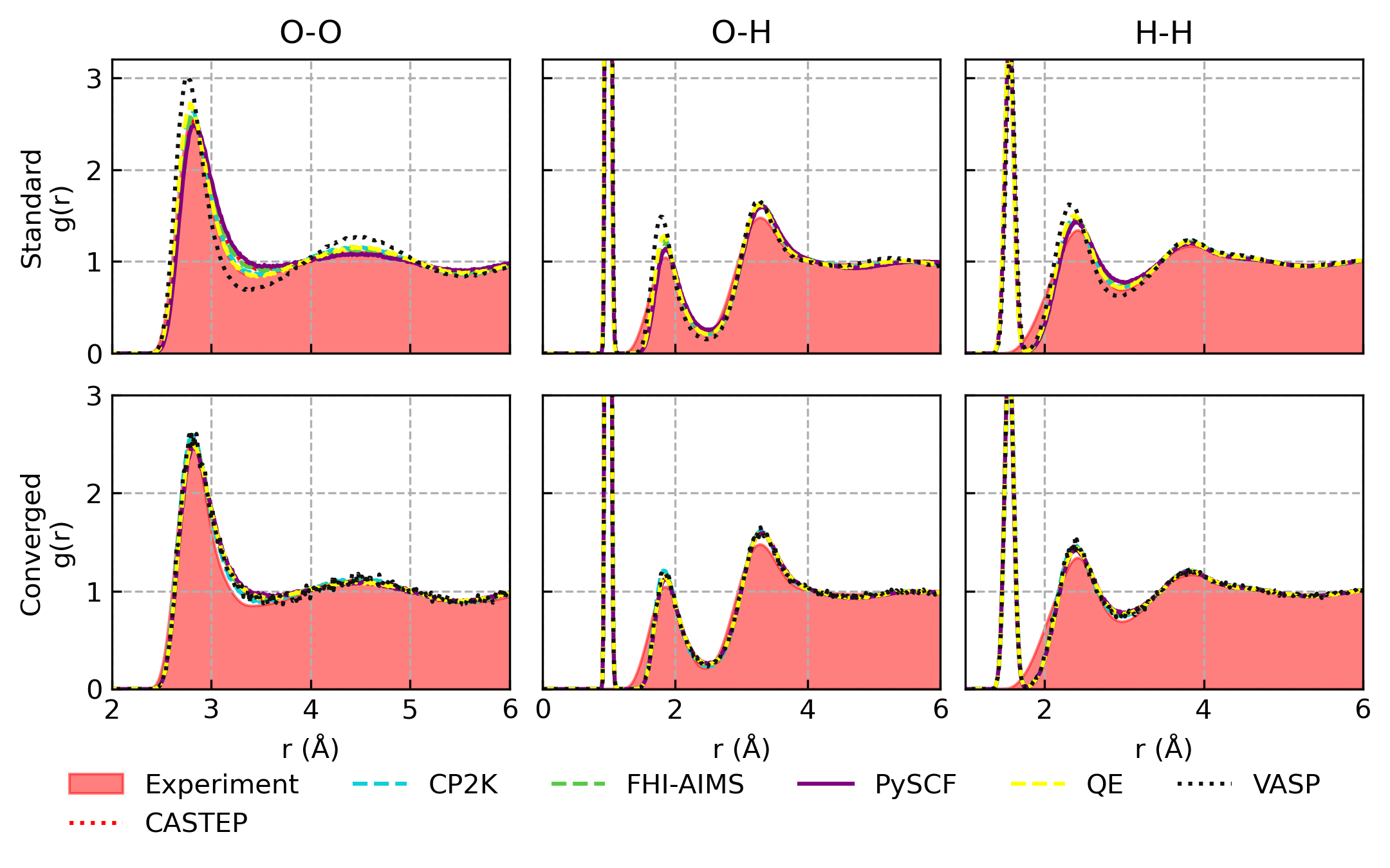}
    \caption{%
    Comparison of O-O, O-H and H-H RDFs predicted by the standard and converged settings of each code. 
    }
    \label{fig:rdfs}
\end{figure}

\subsection{WATER27 interaction energies}

In Table~\ref{tab:water27}, we compare binding energies for the 27 water clusters within the WATER27 dataset predicted by FHI-aims with converged settings against all-electron PySCF with converged settings (i.e., def2-QZVPPD basis set).
For PySCF, we also include estimates using the GTH-PBE pseudopotential with the aug-QZV2P basis set as well as with the TZV2P basis set.
The FHI-aims estimates are taken to be the reference, to compute mean absolute errors (MAEs) per water molecule, to understand the effect of these approximations.
The deleterious effect of using a TZV2P basis set is clear within the interaction energies, where the MAE per water molecule increases from $0.3\,$kcal/mol for the aug-QZV2P basis set to $1.1\,$kcal/mol for the TZV2P basis set.
However, the effects of using the GTH-PBE pseudopotential is not apparent, as the MAE of $0.3\,$kcal/mol is the same irregardless of whether an all-electron or GTH-PBE pseudopotential calculation is performed.
These results give some indication that the errors in the GTH-PBE pseudopotential cannot be established from just simple $0\,$K energetic benchmarks.

\clearpage

\begin{table}
\caption{\label{tab:water27}Comparison of the interaction energy (kcal/mol per molecule) for the WATER27 dataset~\cite{bryantsevEvaluationB3LYPX3LYP2009,anackerNewAccurateBenchmark2014} between revPBE-D3 with FHI-aims, PySCF using all-electron (AE) and PBE-optimized Goedecker-Teter-Hutter (GTH) pseudopotentials with augmented QZ-quality basis sets, as well as PySCF with GTH pseudopotentials and the TZV2P basis set.}
\begin{tabular}{lrrrrr}
\toprule
Reaction & CCSD(T)~\cite{bryantsevEvaluationB3LYPX3LYP2009,anackerNewAccurateBenchmark2014} & FHI-aims & PySCF AE & PySCF GTH-PBE & PySCF GTH-PBE (TZ) \\
\midrule
1 & 2.5 & 2.4 & 2.3 & 2.4 & 2.6 \\
2 & 5.2 & 4.9 & 4.9 & 5.1 & 5.5 \\
3 & 6.8 & 6.6 & 6.5 & 6.8 & 7.4 \\
4 & 7.2 & 6.9 & 6.8 & 7.1 & 7.7 \\
5 & 7.7 & 7.4 & 7.3 & 7.6 & 8.3 \\
6 & 7.6 & 7.4 & 7.3 & 7.6 & 8.3 \\
7 & 7.5 & 7.3 & 7.2 & 7.5 & 8.1 \\
8 & 7.4 & 7.1 & 7.0 & 7.3 & 7.9 \\
9 & 9.1 & 8.8 & 8.7 & 9.0 & 9.8 \\
10 & 9.1 & 8.8 & 8.7 & 9.0 & 9.8 \\
11 & 9.9 & 9.7 & 9.5 & 9.9 & 10.7 \\
12 & 10.4 & 10.0 & 9.9 & 10.3 & 11.1 \\
13 & 10.4 & 10.0 & 9.9 & 10.3 & 11.2 \\
14 & 10.5 & 10.1 & 10.0 & 10.4 & 11.3 \\
15 & 16.9 & 17.7 & 17.6 & 17.8 & 18.2 \\
16 & 19.0 & 19.4 & 19.2 & 19.4 & 20.0 \\
17 & 19.2 & 19.2 & 19.1 & 19.3 & 20.0 \\
18 & 16.8 & 16.8 & 16.7 & 17.0 & 17.7 \\
19 & 16.4 & 16.4 & 16.3 & 16.5 & 17.2 \\
20 & 13.3 & 14.6 & 13.3 & 13.8 & 16.5 \\
21 & 16.2 & 16.8 & 15.7 & 16.2 & 18.8 \\
22 & 16.9 & 17.3 & 16.4 & 16.8 & 19.2 \\
23 & 16.9 & 17.1 & 16.3 & 16.7 & 19.0 \\
24 & 17.0 & 17.3 & 16.5 & 16.9 & 19.3 \\
25 & 16.8 & 17.0 & 16.3 & 16.7 & 18.9 \\
26 & 16.5 & 16.7 & 16.0 & 16.5 & 18.5 \\
27 & 29.6 & 23.5 & 23.6 & 22.9 & 23.8 \\ \hline
MAE[FHI-aims] & 0.5 & 0.0 & 0.3 & 0.3 & 1.1 \\
\bottomrule
\end{tabular}
\end{table}

\clearpage

\bibliography{references.bib}